\documentclass[preprint,amsmath,nofootinbib,superscriptaddress]{revtex4}
\usepackage{setspace}
\usepackage{graphicx}
\usepackage{bm}
\usepackage{epsfig}
\usepackage{color}
\usepackage{colordvi}
\newcommand{\nn}{\nonumber}

\newcommand{\beq}{\begin{equation}}
\newcommand{\eeq}{\end{equation}}
\newcommand{\bea}{\begin{eqnarray}}
\newcommand{\eea}{\end{eqnarray}}

\begin{document}

\title{Spin(1)Spin(2) Effects in the Motion of Inspiralling Compact Binaries  at Third Order in the Post-Newtonian Expansion}

\author{Rafael A. Porto} 
\affiliation{Department of Physics, University of California Santa Barbara, CA 93106}
\author{Ira Z. Rothstein}
\affiliation{Department of Physics, Carnegie Mellon University,
Pittsburgh, PA 15213}
\begin{abstract}
We use effective field theory techniques to compute the potentials due to 
 spin--spin and spin--orbit effects, from which the ${\cal O}({\bf S}_1 {\bf S}_2)$ contribution to
 the motion of spinning compact binaries to third Post--Newtonian (PN) order follow. We use a  formalism which allows us to impose the spin supplementarity condition (SSC) in a canonical framework to all orders in the PN expansion. We explicitly show the equivalence with our previous results, obtained using the Newton--Wigner SSC at the level of the action for spin--spin and spin--orbit potentials reported in  arXiv:gr-qc/0604099 and arXiv:0712.2032[gr-qc] respectively.
\end{abstract}

\maketitle

\section{Introduction}

NRGR \cite{NRGR}, an Effective Field Theory (EFT) approach to gravity, has emerged
as a powerful tool to systematically describe the dynamics of finite size objects in General Relativity (GR). It has been utilized to calculate higher order spin corrections in the PN expansion \cite{eih,comment}, dissipative effects for non-spinning \cite{dis1} as well spinning objects \cite{dis2},  
radiation reaction effects in the extremal limit \cite{chad} and corrections to thermodynamic
quantities in caged black holes \cite{cbh,kol}. In this paper we will extend the formalism for spin in NRGR originally developed in \cite{nrgr3}. In particular we will demonstrate how to calculate the equations of motion (EOM) using the Routhian formalism discussed in \cite{nrgr5}.  
The leading order (LO) spin--spin and spin--orbit potentials were shown in \cite{nrgr3}  to reproduce known results \cite{spin,will,will2,kidd} within the Newton--Wigner (NW) and covariant SSCs. In \cite{eih},  the previously uncalculated 3PN spin--spin potential was obtained using the NW SSC at the level of the action. In \cite{nrgr5}, it was argued that within this approach,  the Hamiltonian method is accurate up to 4PN in the ${\bf S}_1{\bf S}_2$ sector, when curvature effects start to play a role, and the canonical structure in the reduced phase space, $({\bf x},{\cal P},{\bf S})$, is modified. However, in order to calculate the ${\cal O}({\bf S}_1 {\bf S}_2)$ contributions to the EOM at 3PN, the spin--spin potential in \cite{eih} is not sufficient. This fact was made clear by an independent calculation in \cite{Schafer3pn} where the complete potential was computed using a more traditional methodology. Within our approach, we must also include a correction stemming from a subleading effect due to spin--orbit interactions \cite{comment}. Once this extra piece is included, the results in \cite{eih, comment} and \cite{Schafer3pn} agree. As we will see, the Routhian formalism provides yet another independent cross check of the new results to 3PN. 

While working within the NW SSC at the level of the action is relatively simple for LO effects, and ${\cal O}({\bf S}_1{\bf S}_2)$ corrections up to 4PN, calculating subleading ${\cal O}({\bf S}_q)$,  or ${\cal O}({\bf S}^2_q)$ effects can be cumbersome within this methodology. Recall that  the NW SSC leads to a canonical structure in the reduced phase space only in a flat spacetime background \cite{nrgr5}. For ${\cal O}({\bf S}_q)$ effects this structure is lost already at 2.5PN. It is thus desirable to have a technique where the SSC is not imposed until the end of the calculation thus avoiding complicated algebraic structures. Here we elaborate upon such an  approach, presented in  \cite{nrgr5},  and compute the spin--spin potential to 3PN in the covariant SSC.  

Within the NRGR formalism spin--spin, or spin--orbit, refer to the type of diagrams contributing to the potential \cite{nrgr3,eih}. Since we will postpone the SSC to the later stages of the computation, our results for the potentials and EOM will be written in terms of the spin tensor in the local frame, e.g. $S^{ab}$. Therefore, a spin--orbit term proportional to $S^{j0}$ in the EOM can contribute at ${\cal O}({\bf S}_1{\bf S}_2)$ once the SSC is enforced, the spin tensor is reduced to a three vector and the velocity in the local frame is transformed to the global PN frame. Furthermore,  we will show that the spin--spin potential to 3PN also depends on $S^{j0}$, and therefore it will contribute at higher orders in the ${\bf S}^2_1{\bf S}_2$ and ${\bf S}_1{\bf S}_2^2$ sectors.\\

Here we will present some of the details of the calculation of the 3PN potential, as well as including the EOM for the spin of the constituents.  We will first calculate the LO EOM due to the spin--orbit potential and show that we reproduce the well known results before moving on to the spin--spin 3PN computation. For completeness, finally we show the equivalence with our previous results using the NW SSC at the level of the action of \cite{eih,comment}, by constructing an effective potential which agrees with the results in \cite{eih,comment}, and from which the EOM follow via the canonical methods, thereby providing a formal proof of the claims in \cite{eih,comment, nrgr5}.\\

\section{Review of Spin in GR}
The extension of NRGR to include spin effects was
achieved in \cite{nrgr3} by adding world--line degrees of freedom $\Lambda_a^J(\lambda)$, which is the boost that transforms the locally flat frame (labelled by small Roman letters),  to the co-rotating frame labelled by capital Roman letters. The generalized angular velocity is given by
$\Omega^{\mu\nu}=e^{\mu J}\frac{De^\nu_J}{d\lambda}$, where $e^\mu_I= e^\mu_a \Lambda^a_I$ and
$e^\mu_a$ are the co--rotating and locally flat basis respectively (verbeins) and  $e^a_{\mu}e^b_{\nu} g^{\mu\nu} = \eta^{ab}$.
The spin $S_{\mu\nu}$ is introduced as the conjugate momentum to $\Omega_{\mu \nu}$.  The form of
the world--line action is then fixed by reparametrization
invariance \cite{eih},
\begin{equation} \label{action}
S=-\sum_i \left(\int p^\mu_i u^i_{\mu}d\lambda_i + \int
\frac{1}{2}S_i^{IJ}\Omega^i_{IJ} d\lambda_i\right),
\end{equation}
where the sum extends over the consituents , and $S^{IJ} \equiv S^{\mu\nu}e^I_\mu e^J_\nu$. Here we have not included higher dimensional operators which account for finite size effects.  Corrections due to finite size effects are reported in \cite{nos} . The Mathisson--Papapetrou (MP) equations \cite{papa} follow from (\ref{action}) \cite{nrgr3,eih}. The spin--gravity coupling in (\ref{action}) can be rewritten by introducing the Ricci rotation coefficients, $\omega_\mu^{ab} = e^b_\nu D_\mu e^{a\nu}$, as \cite{nrgr3,nrgr5}
\begin{equation} \label{act2}
S_{spin-gravity} =  -\frac{1}{2} \int S_{Lab}\omega^{ab}_\mu u^\mu
d\lambda,
\end{equation}
with $S_L^{ab} \equiv S^{\mu\nu}e_{\mu}^a e_{\nu}^b$, the spin in the locally flat frame (we drop the 
$L$ from now on). By further expanding (\ref{act2}) in the weak gravity limit one obtains the Feynman rules \cite{nrgr3,eih}. Let us emphasize that the SSC is imposed in the local  frame.

\section{The Routhian approach for spinning bodies in NRGR}

A Routhian formalism\footnote{A similar Routhian was originally proposed in \cite{yee} with $S^{ab}u_b=0$ as SSC, which is equivalent to ours at 3PN. See appendix A.} was introduced within the covariant SSC in \cite{nrgr5}.  In what follows we will adopt this framework and compute the 3PN corrections to the potential.  
The virtue of the Routhian formalism is that it allows us to consistently impose, and preserve upon evolution, the SSC in a canonical framework, and properly account for $S^2$ corrections to the potential. The price to pay is that we will work with a spin tensor, $S^{ab}$, rather than a three vector. However, we will show later on that an effective potential in terms of (${\bf x},{\bf v},{\bf S}$) exists, which turns out to be equivalent to our previous results in \cite{eih,comment}.\\

Since the spin is a conjugate momentum, we would like to treat the spin within a Hamiltonian formalism. Whereas, for the worldline position we would like to work within the Lagrangian formalism.  That is, we would like to Legendre transform the Lagrangian with respect to the wordline spin degrees of freedom only. This is done within what is called the ``Routhian" formalism \cite{LL}. We will work in the covariant SSC,
\beq
\label{p.s}
p_a S^{ab} = p_\mu S^{\mu \nu}=0,
\eeq
with $p_\mu$ the coordinate momentum of the particle. To dynamically maintain this conditions we need to impose
\beq
\frac{D}{D\lambda}(p_\mu S^{\mu \nu})=0,
\eeq
and utilizing the MP equations (which follow from (\ref{action}))
\beq
\frac{D S^{\mu\nu}}{D\lambda} 
=p^\mu u^\nu-p^\nu u^\mu , ~~~~~~ \frac{D p^\gamma}{D\lambda}= -\frac{1}{2} R^\gamma_{\rho\alpha\beta}S^{\alpha\beta}u^\rho,
\eeq
yields the momentum
\begin{equation}
p^{\alpha}= \frac{1}{\sqrt{u^2}}\left(m
u^{\alpha}+\frac{1}{2m}R_{\beta\nu\rho\sigma}S^{\alpha\beta}S^{\rho\sigma}u^{\nu}\right)\label{up}.
\end{equation}
Notice that $p \cdot u = m$  once the SSC is
imposed. 

We introduce now the following Routhian  \cite{nrgr5}
\begin{equation} \label{actR}
{\cal R} =-\sum_i \left( m_i \sqrt{u^2_i} + 
\frac{1}{2}S_{i}^{ab}\omega_{ab\mu} u^\mu_i
+\frac{1}{2m_i}R_{d
e a b}(x_i)S^{c d}_{i} S^{a b}_{i} \frac{u^e_i u_{ic}}{\sqrt{u^2}}
+\ldots\right),
\end{equation}
where the ellipses represent curvature  terms necessary to account for the mismatch between $p$ and $u$ in (\ref{p.s}). These terms  contribute 
 beyond the  3PN order we work in this paper\footnote{ In other words, to our level of accuracy, we can consider $S^{ab}u_b = 0$.}. In addition there are finite size corrections to (\ref{up}) which are not shown in (\ref{actR})  but can be consistently included (for details  see appendix A) when going to higher orders in the  PN expansion. 
 
 The EOM follow from
\begin{equation}
\frac{\delta }{\delta x^\mu}\int {\cal R} d\lambda=0, \;\;\; \frac{d
S^{ab}}{d\lambda} = \{S^{ab},{\cal R}\}\label{eom1},
\end{equation}
where the algebra for the phase space variables
$(x^\mu,p^\nu,S^{ab})$ is given by
\begin{eqnarray}
\{x^\mu ,{\cal P}_\alpha\}&=& \delta^\mu_\alpha,\;\;\; \{x^\mu,p_\alpha\}=\delta^\mu_\alpha,  \;\;\; \{{\cal P}^\alpha,{\cal P}^\beta\}= 0, \\
 \{x^\mu,x^\nu\} &=& 0, \;\;\; \{p^\alpha,p^\beta\} = \frac{1}{2}{R^{\alpha\beta}}_{ab}S^{ab}, \label{pp}\\
\{x^\mu ,S^{ab}\}&=& 0, \;\;\; \{p_\alpha,S^{ab}\}= \omega^{c[a}_\alpha S^{b]c} , \;\;\; \{{\cal P}^\alpha,S^{ab}\}= 0 \label{ps} \\
\{S^{ab},S^{cd}\} &=& \eta^{ac} S^{bd}
+\eta^{bd}S^{ac}-\eta^{ad} S^{bc}-\eta^{bc}
S^{ad} \label{als}
\end{eqnarray}
with $p^\mu$ related to the canonical momentum by $ {\cal
P}^\mu = p^\mu + \frac{1}{2}\omega^\mu_{ab} S^{ab}$. It can be shown that the EOM are equivalent to the MP equations and that the extra term in (\ref{actR}) guarantees the preservation  of the covariant SSC.
Notice that with our choice of metric convention $(+,-,-,-)$,  the spin vector algebra differs from the canonical SO(3) algebra by a minus sign. We have compensated for this convention choice by the overall minus sign in the Routhian of (\ref{actR}) which allows us to treat ${\cal R}$ as the usual Lagrangian and keep the spinless feynman rules untouched. Therefore, the relationship between the potential and the Routhian stays as before for the Lagrangian, namely \beq V = - {\cal R}, \eeq 
and therefore the spin EOM in terms of the potential take the form,
\beq
\frac{dS^{ab}}{d\lambda} = \{V, S^{ab}\}\label{eomV}.
\eeq

In practice the EOM for spin can be derived from
\beq
\label{trads}
\frac{d {\bf S}}{d t} = \frac{\partial V}{\partial {\bf S}} \times {\bf S}
\eeq
as one would expect, plus corrections from the $S^{0i}$ components. We will study an example in detail later on.
  
According to the program developed in \cite{NRGR}, to calculate the potential we first need to
generate a set of Feynman rules. The potential will then follow by including the appropriate
set of Feynman diagrams.  Once we have the potential in terms of the spin, position and velocities of the binary constituents, we can calculate the EOM using (\ref{eomV}). 

\subsection{The effective action}

Let us elaborate upon the manipulations leading to the potential. 
The EFT approach is built to separate physics at different scales.
Given that the radiation and potential modes have a ratio of  wavelengths of order $v$, 
this allows us to cleanly separate the physics of radiation from that of potentials in
a systematic fashion. This is not to say that potential modes have no effect on radiation.
Indeed, the tail effect arises from the coupling of a radiation graviton to a potential graviton, and
the EFT reproduces known results \cite{unpublished}. The same can be said for the so--called ``memory effect". However, if we are interested in pure potentials we may completely ignore
the radiation mode in the effective action\footnote{The LO radiation effects were computed for spinless and spinning bodies in \cite{NRGR} and \cite{unp2} respectively, and shown to agree with known results.}. As  discussed in \cite{NRGR}, for spinless objects the effective NRGR action follows from the path integral ($q=1,2$)
\beq
\label{PI} \mbox{exp}\left[i S_{NRGR}[x^i_q]\right] = \int {\cal D}H_{\mu\nu} \mbox{exp}\left[i S[H_{\mu\nu},x^i_q] + i S_{GF}\right],
\eeq
which accounts for the vacuum to vacuum amplitude in the presence of sources, in our case the binary.
In the expression above $S=S_{EH}+S_{pp}$, that is the Einstein--Hilbert action plus the wordline sources, and $S_{GF}$ is a suitable gauge fixing term\footnote{The gauge chosen in \cite{NRGR} corresponds to an harmonic condition up to ${\cal O}(G^2)$ corrections (see Eqs. (62)--(65) in \cite{NRGR}).}  \cite{NRGR}. By expanding the Einstein-Hilbert action in the weak gravity limit we can immediately read
off Feynman rules \cite{NRGR}. Once we compute $S_{NRGR}[x^i_q]$ the EOM follow from a minimal action principle \cite{coleman}, since we have yet to perform the path integral over the sources, namely the wordlines. Notice that the kinetic term is a pure phase which factors out of the path integral.  Therefore, by summing Feynman diagrams effectively we are calculating the potential energy \cite{coleman}.\\ 

Within this framework the inclusion of spin is straightforward. The Routhian ${\cal R}$ will replace the worldline $S_{pp}$ action (recall ${\cal R}=-V$) and the path integral in (\ref{PI}) will produce the effective potential, $-V_{NRGR}$, from which the EOM follow via (\ref{eomV}).  Since we are not imposing the SSC until the EOM is obtained, we are always dealing with a canonical structure, although we pay the price of having a spin tensor $S^{ab}$, rather than a three vector. The latter follows once the SSC imposed at the level of the EOM. For the spin dynamics we directly compute the potential and no kinetic piece is necessary. In a sense the spin dynamics has a more direct contact with the usual interpretation of the path integral as providing the energy of the `vacuum' in the presence of the sources. 

Finally note that the extra terms in the action proportional to the SSC effectively act as Lagrange
multipliers, and enforces the conservation of the SSC upon evolution. Given that we are dealing with second class constraints in the SSC, the Lagrange multipliers are  fixed by  preservation of the constraints \cite{teit}. 
Moreover, since the extra term is proportional to the SSC itself, we are
free to use the equations of motion (i.e. perform an implicit coordinate shift)  
 to replace $dp/d\lambda$ using (5) in the Routhian as we do in (\ref{actR}).

\section{An Example: The spin--orbit effects at leading order}

As a warm up let us start by computing the LO effects in the EOM due to the spin--orbit coupling \footnote{Recall that the LO spin--spin potential does not include the troublesome term proportional to $S^{a0}$ \cite{nrgr3}.}. In this section we will quote the contribution from each diagram to the potential. 
The relevant Feynman rules have been relegated to appendix B.
Each diagrams contribution is of the form $ -iV$.

The  LO spin--orbit potential is found by the instantaneous one graviton exchange
diagram, with the LO spin vertex on one world-line and a mass vertex on the other, as discussed in \cite{nrgr3},
\begin{equation}
\label{so15pn}
V^{so}_{1.5PN}= \frac{G_Nm_2}{r^2}n^j\left(S^{j0}_1+S^{jk}_1(v_1^k-2v^k_2)\right)
+ 1 \leftrightarrow 2,
\end{equation}
with $n^j=\frac{(x_1-x_2)^j}{\mid x_1-x_2\mid}$, hence ${\cal R}^{so}_{1.5pn} = -V^{so}_{1.5pn}$.

Applying the algebra (\ref{als}) we find 
\begin{equation}
\frac{dS_{lk}}{dt}= \frac{Gm_2}{r^2}\left[ (n_iv_{1j}-2n_iv_{2j})   (\delta_{li}S_{kj}+\delta_{kj}S_{li}-\delta_{lj}S_{ki}-\delta_{ki}S_{lj} )+(n_{l}S_{k0}-n_{k}S_{l0} 
)\right]. 
\end{equation}
In terms of the spin `vector', $S^l=\frac{1}{2}\epsilon^{ljk}S^{jk}$, we have in the covariant SSC,
\begin{equation}
\frac{d{\bf S} _1}{dt} = 2\frac{m_2
G_N}{r^2}({\bf n} \times{\bf v} )\times{\bf S} _1 +
\frac{m_2 G_N}{r^2}({\bf S} _1\times{\bf n} )\times {\bf v}_1 \label{ds1dt}
\end{equation} 
with ${\bf v} = {\bf v}_1-{\bf v}_2$. The expression in (\ref{ds1dt}) agrees with the known spin precession  \cite{will,kidd} 
\begin{equation}
\frac{d{\tilde {\bf S}} _1}{dt} = \left(2+\frac{3m_2}{2m_1}\right)\frac{\mu
G_N}{r^2}({\bf n} \times{\bf v} )\times{\tilde {\bf S}} _1,  \label{ds1dtn}
\end{equation} 
with $\mu$ the reduced mass, after the transformation \cite{nrgr3,nrgr5}
\begin{equation}
{\tilde {\bf S}} _1 = (1-\frac{1}{2}{\bf \tilde v} _1^2){\bf S} _1 + \frac{1}{2}{\bf \tilde v} _1({\bf \tilde v} _1\cdot{\bf S} _1)\label{pnshift}.
\end{equation}

In the expression of (\ref{pnshift}) ${\bf \tilde v}_1$ is the velocity in the local frame, which agrees with the coordinate veolocity in the PN (global) frame  at LO.
Let us add a few comments about this distinction. First of all, we are dealing with the local spin, therefore to the order we are working at (recall $p^a \sim m u^a +\ldots$), our SSC reads $S^{ab}u_b=0$, with $u^a=e^a_\mu u^\mu$. If  we choose $\lambda=t$, we have $u^\mu \equiv (1,{\bf v})$ and $u^a$ what we denote as $({\tilde v}^0,{\bf \tilde v})$. For the spin--spin dynamics, the relevant (spin-dependent) part of this relationship is (for instance for particle one)
\begin{eqnarray}
{\tilde v}_1^{a=0} &=& 1 + \ldots, \\
{\tilde v}_1^{a=j} &=& v_1^j  +  \frac{G_N}{r^2} S_2^{jk}n^k + \ldots, 
\label{localv}
\end{eqnarray} 
therefore  
\beq
\label{onshell}
S_1^{i0}= S_1^{ij}v_1^j + S_1^{ij}e^j_0({\bf x}_1) +\ldots =({\bf v}_1\times{\bf S}_1)^i  + \frac{G_N}{r^2}\left[({\bf n} \times {\bf S}_2)\times {\bf S}_1\right]^i +\ldots ,
\eeq 
where we used $e^j_0 ({\bf x}_1)= \frac{G_N}{r^2} ({\bf n}\times {\bf S}_2)^j$, which follows from the one point function, $\langle H^j_0\rangle/2$, or simply inspection of the Kerr metric in harmonic coordinates.
This will add an extra piece in the spin--spin EOM from the LO spin--orbit term of (\ref{so15pn}) (see (\ref{eomspin})), since there is modification in the algebra given by
\beq
\{S_1^i,S_1^{j0}\} = \epsilon^{ijk}S_1^{0k}= S_1^iv_1^j-S_1^jv_1^i+\frac{G_N}{r^2}(({\bf n}\times{\bf S}_2)^jS_1^i-({\bf n}\times{\bf S}_2)^iS_1^j), 
\label{modal}
\eeq
which will lead to a 3PN contribution in the potential\footnote {The extra term in (\ref{onshell}) also provides an extra piece in the potential within NW SSC \cite{comment}.}. The expression in (\ref{modal}) is the main reason why we need to keep track of spin--orbit terms in the potential which will wind up contributing at ${\cal O}({\bf S}_1{\bf S}_2)$ in the EOM.\\

\subsection{The Equivalence of Methodologies}

\subsubsection{The PN frame versus the local frame in the covariant SSC}

Naive comparison of the EOM in (\ref{ds1dt}) with the spin EOM in the covariant SSC, for instance in \cite{owen}, shows that they are indeed different expressions. To understand the discrepancy,  we have to transform from the locally flat frame where (\ref{ds1dt}) is defined, to the commonly used PN frame by rotating the spin tensor using the vierbein and the metric at 1PN order \cite{mtw}, 
\begin{equation}
S_1^{ij} = {\bar S}_1^{ij} + {\bar S}_1^{ik} \frac{h^j_k}{2} - {\bar S}_1^{jk}\frac{h^i_k}{2} + \ldots = {\bar S}_1^{ij}+ 2 \frac{G_Nm_2}{r}{\bar S}_1^{ij}+\ldots, \label{spn}
\end{equation}
with ${\bar S}^{ij}$ the spin tensor in the PN frame within the covariant SSC. One can now trace the disagreement back to the definition of the spin vector. In our calculations we introduced $S^{jk} = \epsilon^{jkl}{\bf S}^l$ in the local frame, however, more generally  we may define the spin four vector as
\begin{equation}
{\bar S}^{\mu\nu} = \frac{1}{m\sqrt{g}} \epsilon^{\mu\nu\alpha\beta} p_\beta {\bar S}_\beta . \label{sp}
\end{equation}

Using now (\ref{sp}) (for instance for $S_1$) we have in the PN frame 
\begin{equation}
\label{sp2}
{\bar S}_1^{ij} = \epsilon^{ijk} \left[ \left(1+\frac{{\bf v}_1^2}{2}-\frac{G_Nm_2}{r}\right) {\bar S}_1^k - v^k_1 ({\bf \bar S}_1\cdot {\bf v}_1)\right]+\ldots
\end{equation}

Leaving
\begin{equation}
{\bf S}_1 = \left(1+2\frac{G_Nm_2}{r}\right) \left[ \left(1+\frac{{\bf v}_1^2}{2}-\frac{G_Nm_2}{r}\right) {\bf \bar S}_1 - {\bf v}_1 ({\bf \bar S}_1\cdot {\bf v}_1)\right]+\ldots,
\end{equation}
which we can expand at 1PN order,
\begin{equation}
{\bf S}_1 = \left(1 +\frac{{\bf v}_1^2}{2}+\frac{G_N m_2}{r} \right) {\bf \bar S}_1 -  {\bf  v}_1 ({\bf \bar S}_1\cdot {\bf v}_1)+\ldots
\end{equation}

The EOM in terms of ${\bf \bar S}$ reads (at 1.5PN),
\begin{equation}
\frac{d{\bf \bar S} _1}{dt} =\frac{m_2
G_N}{r^2}\left[  {\bf \bar S}_1 ({\bf n} \cdot {\bf v}) - 2 {\bf n}({\bf \bar S} _1\cdot{\bf v}) +  ({\bf \bar S}_1 \cdot {\bf n}) ({\bf v}_1-2{\bf v}_2) \right], 
\label{pns}
\end{equation} 
which agrees with the results in \cite{owen, buon}.\\

\subsubsection{Imposing the SSC before or after calculating the EOM}

It is also instructive to see how applying the SSC prior to finding the EOM leads to the same results as in the Routhian approach, where the SSC is enforced after the EOM is obtained. In  \cite{nrgr3} it was shown how the spin EOM in the covariant SSC follow from (\ref{so15pn}) once the SSC is imposed and the non canonical algebra taken into account.  
Let us see how this works for the EOM arising from the  the spin--orbit interaction. Recall the commutators after imposing the covariant SSC are ($Db$ stands for Dirac bracket) \cite{regge}
\begin{eqnarray}
{[x^i_q,x^j_q]}_{Db}&=&\frac{S_q^{ij}}{m_q^2} \label{dirac1}\\
{[x^k_q,S_q^{ij}]}_{Db}&=&\frac{1}{m_q}(S_q^{ik}v^j_q-S_q^{jk}v^i_q)\label{dirac2},\\
 {[S_q^{ab},S_q^{cd}]}_{Db} &=& (\eta^{ac}-\frac{u_q^au_q^c}{u_q^2}) S_q^{bd} +(\eta^{bd}-\frac{u_q^bu_q^d}{u_q^2}) S_q^{ac}
\nn \\& & -(\eta^{ad}-\frac{u_q^au_q^d}{u_q^2}) S_q^{bc}-(\eta^{bc}-\frac{u_q^bu_q^c}{u_q^2})
S^{ad}_q\label{dirac3}
\end{eqnarray}
with $q=1,2$.
(\ref{dirac1})  contributes a non-canoncial  piece to the acceleration
\begin{equation}
\label{newt}
\delta {\bf a}_1= \frac{d}{dt}\left(\left[{\bf x}_1,-\frac{G_Nm_2}{r}\right]_{Db}\right)= 
G_N\frac{m_2}{m_1}\frac{d}{dt} \left(\frac{{\bf n} \times
{\bf S} _1}{r^2}\right)+\ldots
\end{equation}
  In this equation we have left off the other contributions in the RHS.

Within the Routhian approach we are advocating here it is simple to show how (\ref{newt}) arises. First of all we re--write (\ref{so15pn}) as
\begin{equation}
\label{so215pn}
V^{so}_{1.5pn} = \frac{G_Nm_2}{r^2}n^j\left(S_1^{jl}v_1^l+ S^{jk}_1(v_1^k-2v^k_2)\right)
+ \frac{G_Nm_2}{r^2}n^j(S^{j0}_1- S_1^{jl}v_1^l)+ 1 \leftrightarrow 2.
\end{equation}
The second piece would  have vanished had we imposed the covariant SSC and  is  thus responsible 
for the new contribution on the RHS of (\ref{newt}).  It is clear that the only term which does not cancel out once the covariant SSC is imposed, is the one coming from $\frac{d}{dt}\frac{\partial V^{so}_{1.5pn}}{\partial v_1^l}$. That contributes precisely the extra term in (\ref{newt}). The resulting EOM reads (at 1.5PN)
\begin{eqnarray}
 {\bf a}_1^{so}   &=& \frac{G_N}{r^3} \left\{ \frac{m_2}{m_1} \left[ -3{\bf v} \times {\bf S}_1 + 6 {\bf n} ({\bf v} \times {\bf S}_1)\cdot {\bf n} + 3 {\bf n}\cdot{\bf v} ({\bf n} \times {\bf S}_1) \right] \right. \nn \\
& & \left. -4 {\bf v} \times {\bf S}_2 + 6 {\bf n} ({\bf v} \times {\bf S}_2)\cdot {\bf n} + 6 {\bf n}\cdot{\bf v} ({\bf n} \times {\bf S}_2)\right\} \label{covaso}.
\end{eqnarray}

To establish the full equivalence we need to show that  the spin equations also match.  If we impose the covariant SSC in (\ref{so215pn}) and 
use Hamilton's equations, including the correction due to (\ref{dirac2}), we can show that the new piece due to the algebra is (for particle one)
\begin{equation}
\delta \dot {\bf S}_1 = \frac{m_2 G_N}{ r^2}({\bf S} _1\times{\bf n} )\times {\bf v}_1,
\end{equation}
and the equivalence is thus proven. For more on the consistency of the Routhian approach see the Appendix.\\ 

Another important point in the connection between the NW and covariant SSCs is that in addition to (\ref{pnshift}) it also entails a coordinate transformation given by \cite{kidd,regge}
\beq
{\bf x}_q \to {\bf x}_q - \frac{1}{2m_q}({\bf v}_q\times {\bf \tilde S}_q) +\ldots , \\
\label{pnshift2}
\eeq
for $q=1,2$. Equivalently we have
\beq
{\bf r} \to {\bf r} - \frac{1}{2M}({\bf v}\times { \tilde \xi}),
\label{pnshiftv}
\eeq
with ${\bf \xi} = \frac{m_2}{m_1}{\bf S}_1+ \frac{m_1}{m_2}{\bf S}_2$.
This transformation, implemented in the LO EOM, allows us to transform the acceleration from the covariant ( \ref{covaso} ) to the NW SSC. 

\section{The 3PN spin--spin Potential}

Let us now consider the 3PN potential. The result in \cite{eih} was presented with the NW SSC imposed at the level of the action. Here we will derive in more detail the full expression for the potential in terms of the spin tensor, before imposing the covariant SSC. The resulting potential reproduces that of \cite{eih} once the NW SSC is enforced. However, we will retain the expression in terms of $S^{ab}$ and obtain the EOM via (\ref{eomV}). Only then will we impose the covariant SSC. As we will show later on, the EOM obtained with either procedure are equivalent in the ${\bf S}_1{\bf S}_2$ sector up to 4PN order as originally argued in \cite{eih,nrgr5}.

Following the usual rules we draw all possible Feynman diagrams which scale as $v^6$. Each one of this diagrams will contribute to the effective potential by the rule $-i \int V dt  = \mbox{diagram}$ \cite{NRGR}, where only connected diagrams contribute. For simplicity in what follows we will suppress the factors of $\int dt$.\\ 

\begin{figure}[h!]
   \centering
    \includegraphics[width=12cm]{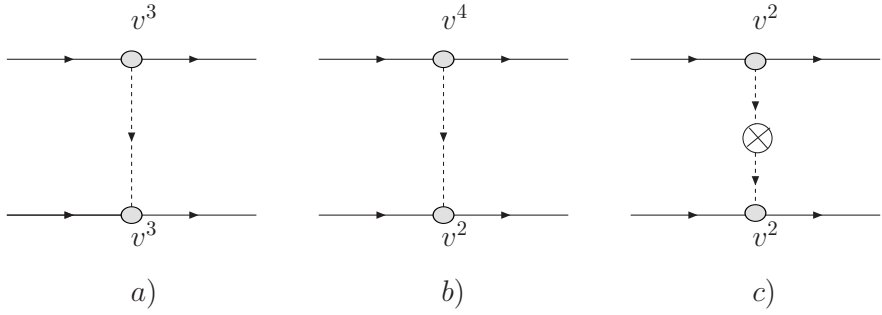}
\caption[1]{ Diagrams contributing to 3PN order which do not
involve non--linear interactions. The blob represents a spin
insertion and the cross corresponds to a propagator
correction.}\label{ss3}
\end{figure}

To calculate the one graviton exchange contribution we should in principle draw all diagrams with propagator which connects to vertices which have subleading scalings. These diagrams are collected in Figs.~\ref{ss3}(a,b,c). This would be formally the correct way to do the calculation in the spirit of effective
field theory, in that each diagram would scale homogeneously. However, in practice it is sometimes
simpler to calculate the full covariant one graviton diagrams and then break it into its individual
pieces which scale homogeneously. This allows us to calculate multiple diagrams simultaneously.
The instantaneous one--graviton exchange can be combined into a single calculation stemming from the linear spin--gravity coupling. If we denote this contribution to the effective potential by $V_{1g}$, and include also the LO piece,  the combination of diagrams then reads  
\begin{equation}
-iV_{1g}=\left(  \frac{i}{2 m_p^2}\right)^2    S^{\mu \beta}_1 u^\alpha_1   S^{\lambda \rho}_2 u^\sigma_2  \langle H_{\alpha\mu,\beta}(x_1)H_{\sigma\lambda,\rho}(x_2) \rangle
\end{equation}
where $\langle ~ \rangle$ represents the Wick contraction.

Let us start by considering the LO contribution from the instantaneous propagator which comes from the  spatial components $c=i, d=j$, since temporal derivatives are down by $v$  and $S_{0i}$ is down by a factor of v. The result reads
\begin{equation}
-iV^{LO}_{1g}=\left(  \frac{i}{2 m_p^2}\right)^2    S^{ik}_1   S^{lj}_2   \langle H_{0i,k}(x_1)H_{0l,j}(x_2) \rangle
\end{equation}
with 
\begin{equation}
\partial^{x_1}_i \partial^{x_2}_j \frac{1}{|{\bf x}_1-{\bf x}_2|} \equiv \partial_{ij} \frac{1}{r} = \frac{1}{r^3}\left(\delta_{ij} - 3 n_i n_j \right).  
\end{equation}

\begin{figure}[h!]
   \centering
    \includegraphics[width=4cm]{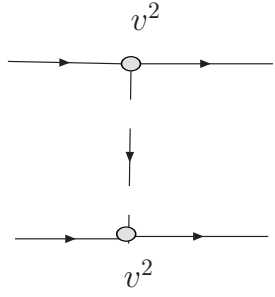}
\caption[1]{ Leading order spin--spin interaction.}\label{sslo}
\end{figure}

From here we read off the LO spin--spin potential (see Fig. \ref{sslo})
\begin{equation}
V^{s1s2}_{2PN}=-\frac{G_N}{r^3}\left({\bf S}_1\cdot {\bf S}_2-3\frac{{\bf r}
\cdot {\bf S}_1 {\bf r} \cdot {\bf S}_2}{r^2}\right). \label{Ess} \end{equation}

Now let us consider the subleading contributions. There are multiple terms at 3PN.
A factor of $v$ arises from either a spatio-temporal component  $S_{0i}$, a temporal derivative, or an
explicit factor of $v$.  Expanding out and keeping only the terms which contribute at 3PN gives
\bea
& &\left(\frac{i}{2M_{pl}}\right)^2 \left[  S^{0i}_1 S^{0j}_2\langle H_{00,i}H_{00,j}\rangle+  S^{ij}_1 S^{nm}_2 v^k_1v^l_2\langle H_{ki,j}H_{ln,m}\rangle + \left( S^{ij}_1 S^{0m}_2 v^k_1\langle H_{ki,j}H_{00,m}
\rangle  \right. \right. \nn \\
 & & + \left. \left. S^{i0}_1 S^{lk}_2\langle H_{0i,0}H_{0l,k}\rangle
+S^{0i}_1 S^{mn}_2v^k_1\langle H_{k0,i}H_{0m,n}\rangle + 1 \leftrightarrow 2 \right) \right] \eea

\begin{figure}[h!]
    \centering
    \includegraphics[width=10cm]{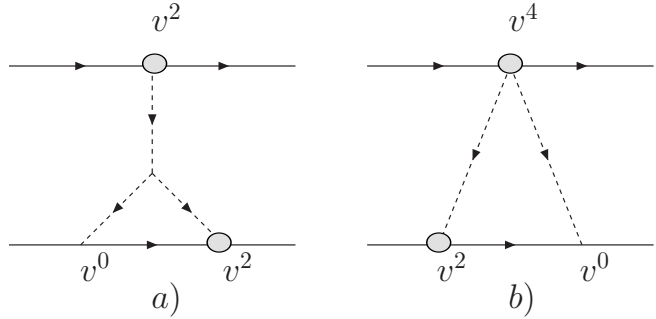}
\caption[1]{Non--linear contributions to the 3PN spin--spin potential.}\label{ss2}
\end{figure}

The evaluation of these integrals is straightforward. Here we evaluate  one particular integal which needs more delicate consideration.
Namely the contribution where one $v$  comes  from a temporal derivative while the other
comes from $S^{0i}$. 
This contribution is given by
\bea
 S^{i0}_1S^{lk}_2 \langle H_{0i,0}(x_1)H_{0l,k}(x_2)\rangle
&=& -\frac{i}{2}S^{i0}_1S^{ik}_2  \int \frac{[d^4p]}{p^2}(\partial^1_0e^{-ip_0(t_1(\lambda_1)-t_2(\lambda_2))})(\partial^2_k e^{-i{\bf p} \cdot ({\bf x}_1(\lambda_1)-{\bf x}_2(\lambda_2))} )\nn \\
&=& -  \frac{i}{2}S^{i0}_1S^{ik}_2  \int \frac{[d^4p]}{{\bf p}^2}(\partial^2_0e^{-ip_0(t_1(\lambda_1)-t_2(\lambda_2))})(\partial^2_k e^{-i{\bf p} \cdot ({\bf x}_1(\lambda_1)-{\bf x}_2(\lambda_2))} )\nn \\
\eea
Now recall  that $\frac{d S^i}{dt} \sim v^2 \frac{v}{r} S^i$ so  we can neglect the time variation of spin at the order we are working at \cite{nrgr3}. It is therefore convenient to trade $\partial_{t_1}$ for $\partial_{t_2}$ picking up a minus sign, and integrating by parts with no net effect. Had we kept $\partial_{t_1}$ we would have to deal with $\frac{d}{dt} S^{0i}$ which we can not neglect. To see this notice that,  imposing  the SSC  would introduce an acceleration dependent piece into the Lagrangian, which can be eliminated using the LO EOM.  However doing so entails a change of coordinates \cite{sch} which is not preferable\footnote{Incidentally, had we insisted on keeping $\partial_{t_1}$ and included this acceleration piece, it turns out that imposing the NW SSC at the level of the action would reproduce the exact same form for the 3PN Hamiltonian of \cite{Schafer3pn}. That is actually the case due to a cancelation of this ${\cal O}({G_N^2})$ acceleration piece with the extra term stemming from subleading corrections in the spin--orbit potential due to (\ref{modal}). A similar result can be found in \cite{levi}. Within the Routhian approach the trade for $\partial_{t_2}$ is preferable.} . If we consider now the instantaneous interaction once again we get
\bea
 S^{i0}_1S^{lk}_2 \langle H_{0i,0}(x_1)H_{0l,k}(x_2)\rangle
&=&  \delta(t_1-t_2)   \frac{i}{2}S^{i0}_1S^{ik}_2 \partial^2_k \partial^2_0  \int \frac{[d^3p]}{{\bf p}^2}(e^{-i{\bf p} \cdot ({\bf x}_1(\lambda_1)- {\bf x}_2(\lambda_2))} )\nn \\
&=& \delta(t_1-t_2)   \frac{i}{8\pi r^3}S^{i0}_1S^{ik}_2(3n^k {\bf n}\cdot {\bf v}_2-v_2^k).
\eea

The net result from the instantaneous one graviton exchange is then given by
\bea
-iV^{inst}_{1g}&=&\frac{iG}{r^3}(\delta^{ij}-3n^i n^j)\left[ S^{i0}_1S^{j0}_2+{\bf v}_1 \cdot {\bf v}_2 
S^{in}_1S^{jn}_2+v_1^mv_2^k S^{ik}_1S^{jm}_2\right.  \\
& -& \left. v_1^k v_2^m S^{ik}_1S^{jm}_2+
S^{i0}_1S^{jk}_2(v_2^k-v_1^k)+S^{ik}_1S^{j0}_2(v_1^k-v_2^k)\right. \nonumber \\
& & + \left. (3n^l{\bf v}_2 \cdot{\bf n} -v_2^l) S_1^{0k}S_2^{kl} + (3n^l{\bf v}_1 \cdot{\bf n} -v_1^l) S_2^{0k}S_1^{kl}  \right]
\eea

The first corrections to instantaneity comes from the diagram shown in  Fig.~\ref{ss3}c.
This correction comes from expanding the graviton propapagtor to second order in the
energy which is suppressed by a factor of $v$ relative to the spatial momentum, i.e
\beq
\frac{1}{p_0^2-{\bf p}^2} \approx -\frac{1}{{\bf p}^2}+\frac{p_0^2}{{\bf p}^4}+...
\eeq
 The result from this diagram is very similar to the spinless case and we have
\begin{eqnarray}
& & \mbox{Fig~\ref{ss3}c} = -i\frac{G_N}{2r^3} S_1^{ki}S_2^{kj}\left[{ \bf v} _1\cdot { \bf v} _2(\delta^{ij} - 3 n^i n^j) -3 { \bf v} _1\cdot{\bf n}  { \bf v} _2\cdot{\bf n}  (\delta^{ij}-5n^in^j)\right. \nonumber \\ & & \left. -3{\bf v} _2\cdot{ \bf n} (v_1^jn^i +v_1^in^j) - 3 {\bf v} _1\cdot{\bf n} (v_2^jn^i+v_2^in^j)+v_1^iv_2^j+v_2^iv_1^j\right] 
\end{eqnarray}

Let us now consider  the terms involving non--linear graviton interactions, as shown in  Figs. \ref{ss2}a and \ref{ss2}b. Let us start with the so called ``seagull" topology
\begin{equation}
\mbox{Fig~ \ref{ss2}b} = \frac{i m_1}{16m_p^2} S_1^{lm}S_2^{ij}\langle T( H_{0l,m}(x_1) H_{00}(x_1)( H^{\lambda}_j H_{0\lambda,i}+ H_{0i,k}H_{kj}))\rangle.
\end{equation}

Note that for this diagram there are a few Wick contractions, that is, the two graviton
vertex can contract in two ways with the mass and spin vertex on the opposing world line.
However many of these contractions vanish since index structures vanish, e.g.
\beq
\langle H_{00}H_{0i}\rangle =  0.
\eeq
The result is given by
\begin{eqnarray}
\mbox{Fig~\ref{ss2}b}=  i  \frac{m_1G_N^2}{r^4} S_1^{jk}S_2^{ij}(5n^kn^i-2\delta^{ki})
\end{eqnarray}

Finally we have the diagram with the three graviton interaction. Again there are mutiple ways of doing the Wick contractions.  As previously mentioned, the best way to handle these contraction is using
a symbolic manipulation program \cite{workshop} where symmetrization is simply handled. The integrals are all variations of the same result used in the one graviton exchange diagram. The result for the diagram is
\begin{eqnarray}
& & \mbox{Fig ~\ref{ss2}a}=  i m_1 \frac{G_N^2}{r^4} S_1^{jk}S_2^{ij}(4n^kn^i-\delta^{ki}) \nonumber \\
\end{eqnarray}
Note that in calculating this diagram we encounter multiple power divergent integrals. These divergences can be absorbed into pure counter-terms, since they just renormalize the mass and possibly other quadrupole moments.  It is simple to see how this occurs in a
diagramatic language. The divergences occur when one of the propagators ending on the
line which has a mass insertion is cancelled by a power of ${\bf k}^2$ arises from the momentum
depedence of the three graviton vertex. One of the lines in the diagrams then contracts to a point.
The resulting diagram looks like an interaction between a self energy (mass correction) and the
spin on the opposite line. The result of this renormalization is that we may simply drop these divergent integrals. As was explained in \cite{NRGR} no {\it physical } logarithmic divergences occur until 5PN order for the case of spinless particles. As it was shown in \cite{nrgr3} that is also the case for spinning bodies and logarithmic divergences due to finite size effects do no show up until ${\cal O} (v^{10})$. This generalizes the so called ``effacement'' of internal structure \cite{damour} to the case of spinning bodies \cite{nrgr3}.
The logarithmic divergences are renormalized by absorption into finite size parameters which present a non trivial renormalization group flow. These are tidally induced effects which in turn do not contribute to the metric solution as it is expected from Birkhoff's theorem. However, there are other types of finite size effects, the so called self--induced effects, which do appear at lower orders as explained in \cite{nrgr3}. This kind of effects are encoded in operators whose coefficients are fixed, like the mass, and can be generated by power law divergences \cite{nrgr3}. For instance in the case of a rotating black hole, finite size corrections appear due to the quadrupole moment of the Kerr spacetime. The coefficient is set by the Kerr metric and it is proportional to $S^2$ \cite{mtw}. The LO corrections (at 2PN) were computed in \cite{nrgr3}, and subleading effects are reported in \cite{nos}.\\ 

Gathering all the pieces together, plus mirror images, we have the complete spin--spin potential to 3PN, prior to  imposing the covariant SSC, 
\begin{eqnarray}
\label{ss312}
V^{spin} &=& -\frac{G_N}{r^3}\left[(\delta^{ij}-3n^i n^j)\left( S^{i0}_1S^{j0}_2+\frac{1}{2}{\bf v}_1 \cdot {\bf v}_2 
S^{ik}_1S^{jk}_2+v_1^mv_2^k S^{ik}_1S^{jm}_2- v_1^k v_2^m S^{ik}_1S^{jm}_2 \right. \right. \nn \\ 
 & & +  \left.
S^{i0}_1S^{jk}_2(v_2^k-v_1^k)+S^{ik}_1S^{j0}_2(v_1^k-v_2^k)\right)+ \frac{1}{2}S_1^{ki}S_2^{kj}\left( 3 { \bf v} _1\cdot{\bf n}  { \bf v} _2\cdot{\bf n}  (\delta^{ij}-5n^in^j) \right. \nn\\   & & + \left. 
   3 {\bf v} _1\cdot{\bf n} (v_2^jn^i+v_2^in^j)+3{\bf v} _2\cdot{ \bf n} (v_1^jn^i +v_1^in^j) -v_1^iv_2^j-v_2^iv_1^j\right) \nn \\ 
 & & + \left. (3n^l{\bf v}_2 \cdot{\bf n} -v_2^l) S_1^{0k}S_2^{kl} + (3n^l{\bf v}_1 \cdot{\bf n} -v_1^l) S_2^{0k}S_1^{kl}  \right] \nn \\ & & +\left(\frac{G_N}{r^3}- \frac{3M G_N^2}{r^4}\right) S_1^{jk}S_2^{ji}(\delta^{ki}-3 n^k n^i)\nn \\
 & & + \frac{G_Nm_2}{r^2}n^j\left(S^{j0}_1+S^{jk}_1(v_1^k-2v^k_2)\right) -\frac{G_Nm_1}{r^2}n^j\left(S^{j0}_2+S^{jk}_2(v_2^k-2v^k_1)\right),
\end{eqnarray}
where we included the LO spin--orbit term which will be relevant latter on due to (\ref{onshell}) and (\ref{modal}). The spin potential in (\ref{ss312}) is the main result of the paper from which the EOM to 3PN order can obtained via (\ref{eomV}).

\subsection{The spin Hamiltonian in the NW SSC to 3PN}

Notice that the spin--spin part of the expression in (\ref{ss312}) agrees with the result reported in \cite{eih} if we impose the NW SSC\footnote{It also provides the spin--orbit potential from which the precession equation follows \cite{nrgr3}.}. However, as we mentioned earlier, to obtain all the contribution in the ${\bf S}_1{\bf S}_2$ sector we need to include subleading corrections in the spin--orbit potential  coming from (\ref{onshell}). The extra term takes the form \cite{comment}
\beq
\frac{G_N}{2r^2}\left( m_2 n^i S_1^{ij}e^j_0({\bf x}_1) - m_1 n^iS_2^{ij}e^j_0({\bf x}_2)\right)= \frac{G^2_NM}{2r^4} \left( ({\bf S}_1\times{\bf n})\cdot ({\bf n}\times{\bf S}_2)\right).
\eeq

For completeness, we present  the ${\bf S}_1{\bf S}_2$ potential in the NW SSC to 3PN order
\begin{eqnarray}
& & V_{NW}^{s1s2} = -\frac{G_N}{2r^3}\left[ {\bf S}_1 \cdot {\bf S}_2\left({3\over2}{\bf v}_1\cdot {\bf v}_2-3{\bf v}_1\cdot {\bf n} {\bf v}_2\cdot {\bf n}
-\left({\bf v}_1^2+{\bf v}_2^2\right)\right)
-{\bf S}_1\cdot {\bf v}_1{\bf S}_2\cdot {\bf v}_2 \nn \right. \\ &-&\frac{3}{2}{\bf S}_1\cdot {\bf v}_2 {\bf S}_2\cdot {\bf v}_1+
{\bf S}_1\cdot {\bf v}_2 {\bf S}_2\cdot {\bf v}_2
+ {\bf S}_2\cdot {\bf v}_1 {\bf S}_1\cdot {\bf v}_1+
3{\bf S}_1\cdot{\bf n}{\bf S}_2\cdot {\bf n}
\left({\bf v}_1\cdot{\bf v}_2+5{\bf v}_1\cdot{\bf n} {\bf v}_2\cdot{\bf n}\right) \nn \\
&-& 3{\bf S}_1\cdot{\bf v}_1{\bf S}_2\cdot {\bf n}{\bf v}_2\cdot {\bf n}-
3 {\bf S}_2\cdot{\bf v}_2{\bf S}_1\cdot{\bf n}{\bf v}_1\cdot{\bf n} + 3({\bf v}_2\times{\bf S}_1)\cdot{\bf n}({\bf v}_2\times {\bf S}_2)\cdot {\bf n}
\nonumber\\
&+& \left.
3( {\bf v}_1\times {\bf S}_1)\cdot {\bf n}( {\bf v}_1\times {\bf S}_2)\cdot{\bf n}-
\frac{3}{2}({\bf v}_1\times{\bf S}_1)\cdot{\bf n}({\bf v}_2\times{\bf S}_2)\cdot{\bf n} - 6({\bf v}_1\times{\bf S}_2)\cdot{\bf n}({\bf v}_2\times{\bf S}_1)\cdot{\bf n}
\right]  \nonumber\\
&+&  
\frac{G^2_N(m_1+m_2)}{2r^4}\left(5{\bf S}_1\cdot{\bf S}_2-17{\bf S}_1\cdot{\bf n}{\bf S}_2\cdot{\bf n}\right) - \frac{G_N}{r^3} \left({\bf S}_1\cdot {\bf S}_2 - 3{\bf S}_1\cdot{\bf n}{\bf S}_2\cdot{\bf n}\right). 
\label{nw3pn}
\end{eqnarray}

As it was argued in \cite{nrgr5} the EOM in the ${\bf S}_1{\bf S}_2$ sector follow from (\ref{nw3pn}) by means of the `traditional' Hamiltonian approach up to 4PN order \cite{comment}. The spin dependent part of the Hamiltonian can be readily obtained from (\ref{nw3pn}), (\ref{Ess}) and (\ref{so15pn}), and takes the form (ignoring 2.5PN spin-orbit and 3PN spin$^2$ terms) to 3PN
\begin{eqnarray}
 H_{NW}^{spin} &=& \frac{G_N}{2m_1m_2r^3}\left[ \frac{3}{2} ({\bf \cal P}_1\times {\bf S}_1)\cdot {\bf n}({\bf \cal P}_2\times {\bf S}_2)\cdot {\bf n} +6 ({\bf \cal P}_2\times {\bf S}_1)\cdot {\bf n}({\bf \cal P}_1\times {\bf S}_2)\cdot {\bf n} \right.  \\ & & -
15  ({\bf \cal P}_1\cdot {\bf n})({\bf \cal P}_2\cdot {\bf n})({\bf S}_1\cdot {\bf n})({\bf S}_2\cdot {\bf n})+\frac{3}{2}  ({\bf \cal P}_2\cdot {\bf S}_1)({\bf \cal P}_1\cdot {\bf S}_2) - \frac{3}{2}  ({\bf \cal P}_2\cdot {\bf \cal P}_1)({\bf S}_1\cdot {\bf S}_2)\nn \\ 
& & - 3({\bf \cal P}_1\cdot {\bf \cal P}_2)({\bf S}_1\cdot {\bf n})({\bf S}_2\cdot {\bf n})+  3({\bf \cal P}_1\cdot {\bf S}_1)({\bf n}\cdot {\bf \cal P}_2)({\bf S}_2\cdot {\bf n})+3({\bf \cal P}_2\cdot {\bf S}_2)({\bf n}\cdot {\bf \cal P}_1)({\bf S}_1\cdot {\bf n})\nn\\
& & \left.  + 3({\bf \cal P}_2\cdot {\bf n})({\bf n}\cdot {\bf \cal P}_1)({\bf S}_1\cdot {\bf S}_2)+({\bf \cal P}_2\cdot {\bf S}_2) ({\bf \cal P}_1\cdot {\bf S}_1)\right]\nn \\ & & +\frac{G_N}{2m_1^2r^3}\left[{\bf \cal P}_1^2({\bf S}_1\cdot {\bf S}_2) - 3 ({\bf \cal P}_1\times {\bf S}_1)\cdot {\bf n}({\bf \cal P}_1\times {\bf S}_2)\cdot {\bf n} - ({\bf \cal P}_1\cdot {\bf S}_2) ({\bf \cal P}_1\cdot {\bf S}_1)\right] \nn\\
& & +\frac{G_N}{2m_2^2r^3}\left[{\bf \cal P}_2^2({\bf S}_1\cdot {\bf S}_2) - 3 ({\bf \cal P}_2\times {\bf S}_1)\cdot {\bf n}({\bf \cal P}_2\times {\bf S}_2)\cdot {\bf n} - ({\bf \cal P}_2\cdot {\bf S}_2) ({\bf \cal P}_2\cdot {\bf S}_1)\right] \nn\\
& & +  
\frac{G^2_N(m_1+m_2)}{2r^4}\left(11{\bf S}_1\cdot{\bf S}_2-23({\bf S}_1\cdot{\bf n})({\bf S}_2\cdot{\bf n})\right) - \frac{G_N}{r^3} \left({\bf S}_1\cdot {\bf S}_2 - 3({\bf S}_1\cdot{\bf n})({\bf S}_2\cdot{\bf n})\right) \nn \\ & & + \frac{G_N}{r^2} \left[ \frac{3m_2}{2m_1} ({\bf n}\times {\bf \cal P}_1)\cdot {\bf S}_1 - 2({\bf n}\times {\bf \cal P}_2)\cdot{\bf S}_1 + 2({\bf n}\times {\bf \cal P}_1)\cdot {\bf S}_2  - \frac{3m_1}{2m_2} ({\bf n}\times {\bf \cal P}_2)\cdot {\bf S}_2 \right]\nn,
\label{nwh3pn}
\end{eqnarray}
where
\beq
{\bf \cal P}_1 = m_1 {\bf v}_1 + 2 \frac{G_N m_1}{r^2}{\bf n}\times {\bf S}_2 + \frac{3 G_N m_2}{2r^2}{\bf n}\times {\bf S}_1,  \:\: 1 \to 2.
\eeq

To obtain the EOM however we will proceed differently, and again we will not impose the SSC up until after we have solved for the EOM resulting from the potential in (\ref{ss312}) using the Routhian approach. As we shall see the extra piece due to spin--orbit effects will come from (\ref{modal}). We will explicitly show however that the results are equivalent. Let us remark that a Hamiltonian similar to that in (\ref{nwh3pn}) was recently found in \cite{Schafer3pn}, and shown to be equivalent in \cite{comment} once the spin-orbit effect is included.

\section{The spin equation of motion to 3PN order}

The 3PN  contribution  to the EOM for spin follows from the potential  in a similar fashion to  the LO spin--orbit example. Let us proceed systematically for particle one. For the spin--spin part of the potential to 3PN in (\ref{ss312}) we have two pieces, one depending on ${\bf S}_1$,  
\begin{eqnarray}
V_{{\bf S}_1} &=& -\frac{G_N}{r^3} \left[ {\bf S}_1\cdot {\bf S}_2 \left( \frac{1}{2} {\bf v}_1\cdot {\bf v}_2 - {\bf v}_2^2-\frac{3}{2}{\bf n}\cdot {\bf v}_1{\bf n}\cdot {\bf v}_2\right) + \frac{3}{2} {\bf n}\cdot {\bf S}_2{\bf n}\cdot {\bf S}_1 \left( {\bf v}_1\cdot {\bf v}_2 +5 {\bf n}\cdot {\bf v}_2{\bf n}\cdot {\bf v}_1\right) \right. \nonumber  \\ & &
+{\bf S}_1\cdot {\bf v}_2 {\bf S}_2\cdot {\bf v}_2 - \frac{1}{2} {\bf S}_1\cdot {\bf v}_1 {\bf S}_2\cdot {\bf v}_2 - \frac{1}{2} {\bf S}_2\cdot {\bf v}_1 {\bf S}_1\cdot {\bf v}_2 -3 {\bf n}\cdot({\bf v}_2\times {\bf S}_1){\bf n}\cdot({\bf v}_1\times {\bf S}_2) \nonumber \\ & & +  3 {\bf n}\cdot({\bf v}_2\times {\bf S}_1){\bf n}\cdot({\bf v}_2\times {\bf S}_2)
+ \frac{3}{2} {\bf n}\cdot{\bf v}_1 {\bf n}\cdot{\bf S}_2{\bf S}_1\cdot{\bf v}_2-\frac{3}{2} {\bf n}\cdot{\bf v}_1 {\bf n}\cdot{\bf S}_1{\bf S}_2\cdot{\bf v}_2  \nonumber \\ & & - \left. \frac{3}{2} {\bf n}\cdot{\bf v}_2 {\bf n}\cdot{\bf S}_1{\bf S}_2\cdot{\bf v}_1 - \frac{3}{2} {\bf n}\cdot{\bf v}_2 {\bf n}\cdot{\bf S}_2{\bf S}_1\cdot{\bf v}_1 \right] \nonumber \\ & & + \left(-\frac{G_N}{r^3}+ \frac{3MG_N^2}{r^4}\right) \left({\bf S}_1\cdot{\bf S}_2 - 3{\bf S}_1\cdot{\bf n}{\bf S}_2\cdot{\bf n}\right),
\end{eqnarray}
and another one 
\begin{eqnarray}
\label{w0}
V_{S^{0i}_1} &=& - A^iS^{i0}_1 \\
{\bf A}_1 &=&  \frac{G_N}{r^3}\left\{(3{\bf v}_2-{\bf v}_1)\times{\bf S}_2 - 3{\bf n}\cdot(2{\bf v}_2-{\bf v}_1)\times{\bf S}_2){\bf n} -3 {\bf n}\cdot{\bf v}_2({\bf n}\times{\bf S}_2)\right\} = \nonumber \\ &=& {\bf \tilde a}^{so}_{1(2)} +\frac{G_N}{r^3}\left(2 {\bf v}_1\times {\bf S}_2-
3  {\bf n}\cdot({\bf v}_1\times {\bf S}_2) {\bf n} -3 {\bf n}\cdot{\bf v}_1({\bf n}\times{\bf S}_2)\right)
\end{eqnarray}
with ${\bf \tilde a}^{so}_{1(2)}$ the ${\bf S}_2$ part of the acceleration in the local frame. 	The latter is given by
\begin{equation}
{\bf \tilde a}^{so}_1 = {\bf a}_1^{so} + \frac{d}{dt} \left( \frac{G_N}{r^2} {\bf n} \times {\bf S}_2\right)+\ldots,  
\end{equation}
 where ${\bf a}_1^{so}$ is the acceleration in the PN frame given in (\ref{covaso}). Then (with $\chi = {\bf S}_2 + \frac{m_2}{m_1}{\bf S}_1$)
\begin{equation}
\label{covasoloc}
{\bf \tilde a}^{so}_1 =  \frac{G_N}{r^3} \left[ -3{\bf v} \times \chi + 6 {\bf n} ({\bf v} \times \chi)\cdot {\bf n} + 3 {\bf n}\cdot{\bf v} ({\bf n} \times \chi) \right]. 
\end{equation}

Notice we also have \beq {\bf A}_1 = {\bf \tilde a}^{so}_{1(2)} + {\bf v}_1\times \omega^{ss}_0,\label{w0A}\eeq with $\omega^{ss}_0$ the LO spin--spin frequency. This expression will be useful later on to prove the equivalence with our previous results in \cite{eih,comment}.\\
 
Using (\ref{eomV}) the ${\cal O}({\bf S}_1{\bf S}_2)$ part of the spin EOM ends up being  
\begin{equation}
\label{eomspin}
\frac{d{\bf S}_1}{dt} = (\omega^{ss}_0+\omega^{ss}_1) \times {\bf S}_1 + ({\bf v}_1\times {\bf S}_1)\times{ \bf A}_1 + \frac{m_2G_N^2}{r^4} {\bf n} \times [ ({\bf n}\times{\bf S}_2)\times{\bf S}_1],
\end{equation}
where the  last term follows from the correction in the spin--orbit part of the potential in (\ref{ss312}) due to (\ref{modal}), and 
\begin{eqnarray}
\omega^{ss}_0 &=& -\frac{G_N}{r^3} \left({\bf S}_2 - 3{\bf n}{\bf S}_2\cdot{\bf n}\right) \\
\omega^{ss}_1 &=& -\frac{G_N}{r^3} \left[ {\bf S}_2\left( \frac{1}{2} {\bf v}_1\cdot {\bf v}_2 - {\bf v}_2^2-\frac{3}{2}{\bf n}\cdot {\bf v}_1{\bf n}\cdot {\bf v}_2\right) + \frac{3}{2} {\bf n} ({\bf n}\cdot {\bf S}_2)\left( {\bf v}_1\cdot {\bf v}_2 +5 {\bf n}\cdot {\bf v}_2{\bf n}\cdot {\bf v}_1\right) \right. \nonumber  \\ & &
+ {\bf v}_2 ({\bf S}_2\cdot {\bf v}_2) - \frac{1}{2}  {\bf v}_1 ({\bf S}_2\cdot {\bf v}_2) - \frac{1}{2}  {\bf v}_2 ({\bf S}_2\cdot {\bf v}_1) -3 ({\bf n}\times {\bf v}_2) {\bf n}\cdot({\bf v}_1\times {\bf S}_2) \nonumber \\ & & +  3 ({\bf n}\times {\bf v}_2){\bf n}\cdot({\bf v}_2\times {\bf S}_2)
+ \frac{3}{2}  {\bf v}_2({\bf n}\cdot{\bf v}_1) ({\bf n}\cdot{\bf S}_2)-\frac{3}{2} {\bf n}({\bf n}\cdot{\bf v}_1) ({\bf S}_2\cdot{\bf v}_2)   \nonumber \\ & & - \left. \frac{3}{2} {\bf n} ({\bf n}\cdot{\bf v}_2) ({\bf S}_2\cdot{\bf v}_1) - \frac{3}{2} {\bf v}_1({\bf n}\cdot{\bf v}_2) ({\bf n}\cdot{\bf S}_2) \right] + \frac{3MG_N^2}{r^4} \left({\bf S}_2 - 3{\bf n}{\bf S}_2\cdot{\bf n}\right)
\label{w1ss}
\end{eqnarray}

In what follows we will show how to reproduce the precession equation and the equivalence with the result of \cite{eih,comment}.

\section{The Precession equation to 3PN and the equivalence with our previous results using the NW SSC}

The precession equation, 
\beq
\label{eomnws}
\frac{d{\bf S}_q}{dt} = \omega^{nw}_q \times {\bf S}_q,
\eeq can be obtained from (\ref{nwh3pn}) with $\omega_q = \frac{\partial H^{spin}_{NW}}{\partial {\bf S}_q}$ ($q=1,2$), for instance for particle 1 (ignoring linear in spin, and also spin$^2$, terms),
\bea
\omega^{nw}_1 &=& 
\label{omegasnw} 
\frac{G_N}{2r^3}\left[ \frac{3}{2} {\bf n}\times {\bf  v}_1({\bf v}_2\times {\bf S}_2)\cdot {\bf n} +6 {\bf n}\times {\bf v}_2({\bf v}_1\times {\bf S}_2)\cdot {\bf n} \right. \nn \\ & & -15 {\bf n} ({\bf v}_1\cdot {\bf n})({\bf v}_2\cdot {\bf n})({\bf S}_2\cdot {\bf n})+\frac{3}{2}  {\bf v}_2({\bf v}_1\cdot {\bf S}_2) - \frac{3}{2}  ({\bf v}_2\cdot {\bf v}_1){\bf S}_2 \nn \\ 
& & - 3 {\bf n}({\bf v}_1\cdot {\bf v}_2)({\bf S}_2\cdot {\bf n})+  3{\bf v}_1({\bf n}\cdot {\bf v}_2)({\bf S}_2\cdot {\bf n})+3 {\bf n}({\bf v}_2\cdot {\bf S}_2)({\bf n}\cdot {\bf v}_1)\nn\\
& & \left.  + 3{\bf S}_2({\bf v}_2\cdot {\bf n})({\bf n}\cdot {\bf v}_1)+{\bf v}_1({\bf v}_2\cdot {\bf S}_2) \right]\nn \\ & & +\frac{G_N}{2r^3}\left[{\bf v}_1^2 {\bf S}_2 - 3 {\bf n}\times {\bf  v}_1 ({\bf v}_1\times {\bf S}_2)\cdot {\bf n} - {\bf v}_1({\bf v}_1\cdot {\bf S}_2) \right] \nn\\
& & +\frac{G_N}{2r^3}\left[{\bf v}_2^2{\bf S}_2 - 3 {\bf n}\times {\bf v}_2({\bf v}_2\times {\bf S}_2)\cdot {\bf n} - {\bf v}_2 ({\bf v}_2\cdot {\bf S}_2) \right] \nn\\
& & +  
\frac{G^2_N(m_1+m_2)}{2r^4}\left(5{\bf S}_2-17{\bf n}({\bf S}_2\cdot{\bf n})\right) - \frac{G_N}{r^3} \left({\bf S}_2 - 3{\bf n}({\bf S}_2\cdot{\bf n})\right). 
\eea

Notice that equivalently we have $\omega_q = \frac{\partial V^{s1s2}_{NW}}{\partial {\bf S}_q}$ ($q=1,2$).\\ 

In what follows we will show  (\ref{eomnws}) is equivalent to (\ref{eomspin}) up to ${\cal O}({\bf S}_1^2)$ effects. To transform the EOM in covariant SSC to NW SSC, as a first step we need to implement the shifts in (\ref{pnshift}) and (\ref{pnshift2}). Recall (\ref{pnshift}) already transforms the LO spin--orbit part of the EOM into a precession equation (see (\ref{ds1dt}) and (\ref{ds1dtn})). Also the coordinate transformation in (\ref{pnshift2}) shifts the form of the frequency in the precession equation from the spin--orbit part. The EOM in terms for ${\bf \tilde S}_1$ reads
\begin{equation}
\label{prewss}
\frac{d{\bf \tilde S}_1}{dt} = {\tilde \omega}^{ss}_1\times {\bf \tilde S}_1
\end{equation}
with
\begin{equation}
\label{eqw1}
{\tilde \omega}^{ss}_1= \delta {\omega}^{so}_1+\omega^{ss}_0+\delta \omega^{ss}_0+ \omega^{ss}_1 + \frac{1}{2}{\bf v}_1 \times {\bf \tilde A}_1 + \frac{1}{2} \frac{m_2G_N^2}{r^4} \left[({\bf \tilde S}_2\times {\bf n})\times {\bf n}\right],
\end{equation} 
\begin{eqnarray} \delta {\bf \tilde \omega}^{so}_1 &=&  \frac{G_N}{2r^3}\left\{{\bf n}\times \left(\frac{9}{2}{\bf v}_1-6{\bf v}_2\right)\left[\frac{m_2}{m_1}({\bf n}\times {\bf v}_1)\cdot {\bf \tilde S}_1 -({\bf n}\times {\bf v}_2)\cdot {\bf \tilde S}_2\right] \right. \nonumber \\ & & \left. + \left( {\bf v}_2\times {\bf \tilde S}_2 -\frac{m_2}{m_1}{\bf v}_1\times {\bf \tilde S}_1\right)\times \left(\frac{3}{2} {\bf v}_1-2{\bf v}_2\right) \right\} \nonumber
\\ & & + \frac{G_N^2m_2}{r^4} {\bf n}\times \left(\frac{3m_2}{4m_1}({\bf n}\times {\bf \tilde S}_1) + \frac{m_1}{m_2}({\bf n}\times {\bf \tilde S}_2) \right),
\label{dw1}
\end{eqnarray}
and
\begin{equation}
\delta \omega^{ss}_0 = -\frac{G_N}{2r^3}\left[ {\bf v}^2_2 {\bf \tilde S}_2 - {\bf v}_2 ({\bf \tilde S}_2\cdot {\bf v}_2)-
3{\bf n} ({\bf n}\cdot {\bf \tilde S}_2) {\bf v}_2^2 + 3 {\bf n} ({\bf n}\cdot {\bf v}_2) ({\bf v}_2\cdot {\bf \tilde S}_2)\right] \label{deltaw0}.
\end{equation}

From the expression in (\ref{dw1}) we will only consider the ${\bf S}_1{\bf S}_2$ contributions.
What we need now is to find an additional contributions (curvature effects) which would take the form of (\ref{eqw1}) into the expression in (\ref{omegasnw}). First of all notice that ${\bf A}_1$ ends up effectively like in the NW SSC, due to the $\frac{1}{2}$ in (\ref{eqw1}). However, there is a piece which differs from the full NW form and comes from the $S^{j0}_2S_1^{i0}$ term in the potential. For the expression in $\omega^{ss}_1$ the difference is just the factor of $\frac{1}{2}$ for $S^{j0}_2$ in the NW SSC. Henceforth, we can split the terms in (\ref{eqw1}) as
\begin{eqnarray}
{\bf \tilde A}_1 &=& {\bf \tilde A}^{nw}_1 - \frac{G_N}{2r^3} \left(3{\bf n} ({\bf n}\times {\bf v}_2)\cdot {\bf \tilde S}_2- {\bf v}_2\times{\bf \tilde S}_2\right) \\
\omega^{ss}_1 &=& {\hat \omega}^{nw}_1 - \frac{G_N}{2r^3} \left[ ({\bf v}_2\times{\bf \tilde S}_2)\times (2{\bf v}_1-{\bf v}_2) 
-3{\bf n}\times({\bf v}_1-{\bf v}_2) ({\bf n}\times {\bf v}_2)\cdot {\bf \tilde S}_2 \right. \nonumber \\ & & \left. + 3{\bf n}\times({\bf v}_2\times{\bf \tilde S}_2)({\bf n} \cdot {\bf v}_1)\right]. 
\end{eqnarray}

Notice that $\omega^{nw}_1 = \frac{1}{2} {\bf v}_1\times {\bf \tilde A}^{nw}_1 + {\hat \omega}^{nw}_1$ and the EOM becomes
\begin{eqnarray}
\frac{d{\bf \tilde S}_1}{dt} &=& \left. \frac{d {\bf S}_1}{dt}\right|_{nw} - \frac{G_N}{2r^3} \left[-3({\bf n}\times {\bf v}_2) ({\bf n}\times {\bf v}_2)\cdot {\bf \tilde S}_2+({\bf v}_2\times{\bf \tilde S}_2)\times \left({\bf v}_1+{\bf v}_2\right)+ {\bf v}^2_2 {\bf \tilde S}_2  \right. \nonumber \\ & & \left. + 3{\bf n}\times({\bf v}_2\times{\bf \tilde S}_2)({\bf n} \cdot {\bf v}_1) -  {\bf v}_2 ({\bf \tilde S}_2\cdot {\bf v}_2) -
3{\bf n} ({\bf n}\cdot {\bf \tilde S}_2) {\bf v}_2^2 + 3 {\bf n} ({\bf n}\cdot {\bf v}_2) ({\bf v}_2\cdot {\bf \tilde S}_2)\right]\times {\bf \tilde S}_1 \nonumber \\ & & + \frac{G_N^2m_1}{2r^4} \left[{\bf n}\times \left({\bf n}\times {\bf \tilde S}_2 \right)\right]\times {\bf \tilde S}_1 
\end{eqnarray}

The extra shift we need to add to (\ref{pnshift}) and transform away the undesired pieces ends up being 
\begin{eqnarray}
{\bf S}_1^{nw}&=& {\bf \tilde S}_1 + \frac{G_N}{2r^2} \left( {\bf \tilde S}_2 ({\bf v}_2\cdot{\bf n})- ({\bf \tilde S}_2\cdot {\bf n}){\bf v}_2\right) \times {\bf \tilde S}_1 +\dots = \\ &=& (1-\frac{1}{2}{\bf \tilde v} _1^2){\bf S} _1 + \frac{1}{2}{\bf \tilde v} _1({\bf \tilde v}\cdot {\bf S}_1)  + \frac{G_N}{2r^2}\left[ ({\bf v}_2 \times {\bf S}_2)\times {\bf n}\right]\times{\bf S}_1+\ldots \label{nlonw},
\end{eqnarray} 
and the equivalence is thus formally proven. Notice that ${\bf S}^{nw}={\bf \tilde S} + {\cal O}(G_N)$. Recall that ${\bf \tilde S}$ reproduces the spin dynamics in NW gauge at LO, in particular the LO precession equation in (\ref{ds1dtn}). However, at next to LO, to transform to the NW gauge we needed to take into account  curvature effects that modify the shift in (\ref{pnshift}). Some of these effects were already included once the velocity in the local frame is transformed to the coordinate velocity in the PN frame (see (\ref{pnshift}) and (\ref{localv})). The extra term, necessary to account for the discrepancy between a flat and curved background, appears in (\ref{nlonw}). Notice that in the limit $G_N \to 0$ these contributions vanish. To avoid confusion, to 3PN one can skip the intermediate step in (\ref{pnshift}) which defines ${\bf \tilde S}$, and use (\ref{nlonw}) to relate ${\bf S}$ with ${\bf S}^{nw}$, the local spin in the covariant and NW  SSC respectively. The equivalence of results thus follows.\\

To show that the position dynamics is also recovered, once the spin EOM is reproduced, we can simply construct an effective potential as
\begin{equation}
V_{eff} =  {\bf \tilde \omega}^{ss}_1\left({\bf S}^{nw}_2\right) \cdot {\bf S}^{nw}_1
\end{equation}
from which the spin corrections to the position dynamics can be derived via the `traditional'  Hamiltonian approach. The above expression obviously reproduces the results of \cite{eih,comment,Schafer3pn}. Nevertheless, the more traditional spin dynamics in covariant SSC is shown in (\ref{eomspin}) with the spin defined in the local frame. To transform to the PN frame one can proceed as we did in (\ref{spn})-(\ref{pns}) for the LO case. 

\subsection{Adding ${\bf S}^2$ terms}

As we mentioned earlier, the spin EOM which follows from (\ref{nw3pn}) fails to  reproduce all of the ${\cal O}({\bf S}_q^2)$ terms. These terms can be computed by working within the Routhian formalism by
  adding  the corrections due to the Riemann dependent term in (\ref{actR}). We may use this term as written or equivalently we may perform a field redefinition such that 
  \beq
\frac{1}{2m_q}R_{d
e a b} S^{c d} S^{a b} \frac{u^e u_{c}}{\sqrt{u^2}} \to \frac{1}{m_q}\frac{Dp_d}{d\lambda}\frac{S^{dc}u_c}{\sqrt{u^2}}\label{dsu}.
\eeq

The procedure for calculating the potential follows the exact same steps as before and it can be shown that \cite{nos} the potential due to this term in the covariant SSC takes the form
\beq
V_{3PN}^{s^2} = \ldots -\left({\bf \tilde a}^{so}_{1(1)}\right)^l S^{lc}_1v_{1c} + 1 \to 2,
\eeq
from which we get the following contribution to  the spin EOM 
\beq
\frac{d{\bf  S}_1}{dt}=
\ldots + ~ ({\bf \tilde a}^{so}_{1(1)}\times {\bf S}_1)\times {\bf v}_1 + \ldots
\eeq
with ${\bf \tilde a}^{so}_{1(1)}$ the ${\bf  S}_1$ dependent part of the spin--orbit acceleration in the local frame (see (\ref{covasoloc})). 
This is however not yet complete since we are still missing $S^2$ corrections stemming from finite size effects, as well as non--linear corrections ($\sim G_N^2$) as in Fig.~\ref{ss2}a, with two LO spin insertions on the same worldline\footnote{Notice that the would be 3PN contribution from a seagull diagram similar to Fig.~\ref{ss2}b, where a non--linear contribution from the mass worldline couples to two spin insertions in the companion worldline, vanishes.}. Finite size effects on the other hand are encoded in higher dimensional operators \cite{NRGR,nrgr3}. For the case of self--induced spin effects the new term in the Routhian takes the form ($q=1,2$)
\begin{equation}
\frac{C^{(q)}_{ES^2}}{2m_q m_p}\frac{E_{ab}}{\sqrt{u_q^2}}
{{\cal S}_q}^a_c {\cal S}_q^{cb},\label{s2} 
\end{equation} 
in the worldline \cite{nrgr3,eih}, where ${\cal S}^{ab}$ is defined as\cite{yee}
\beq
{\cal S}^{ab} = S^{ab} + \frac{u_c}{u^2}S^{c[a}u^{b]},
\eeq
 which guarantees the SSC is preserved in time (see appendix). In the expression of (\ref{s2}) the Wilson coefficients, $C^{(q)}_{ES^2}$, are constants which are determined solely by the nature of the object \cite{NRGR,nrgr3}, and $E_{ab}$ is the electric component of the Weyl tensor in the local frame. In the case of a rotating black hole we have $C_{ES^2}=1$, and this term represents the
non-vanishing quadrupole moment of the Kerr solution. The LO self--induced finite size contribution to the potential thus takes the form \cite{eih,nrgr3}
\beq
V_{2PN}^{s^2} = -C^{(1)}_{ES^2}\frac{m_2}{2m_1r^3} \left({\bf S}_1\cdot{\bf S}_1 - 3{\bf S}_1\cdot{\bf n}{\bf S}_1\cdot{\bf n}\right) + 1\to 2 .
\eeq

Higher order corrections will follow from (\ref{s2}) in a similar manner. We report the full ${\cal O}({\bf S}_q^2)$ contribution in a companion publication \cite{nos}.
 
\section{Conclusions}

In this paper we have presented the details of the calculation of the ${\cal O}({\bf S}_1{\bf S}_2)$ effects to 3PN order. We computed the potential, and showed how to calculate using a Routhian approach, imposing the SSC only at the last stage of the calculation. The EOM follow from (\ref{ss312}) via (\ref{eomV}). We proved the equivalence of this methodology, with a covariant SSC, to that originally espoused in \cite{eih,comment}, where we calculated within the NW imposing the SSC at the level of the action. 
In this paper we have not included effects which go as $S^2$ such as finite size effects. The first non--zero finite size effects for spinless particles start out at 5PN \cite{damour,NRGR}, whereas spin induced finite size effects show up at LO, e.g. 2PN \cite{nrgr3}. Tidally induced finite size effects (logarithmic effects) first appear at 5PN for the case of spinning bodies \cite{nrgr3}.
In a subsequent paper we will present the next to LO $S^2$ effects using the same formalism discussed here. 

\vskip 0.5cm

This work was supported in part by the Department of Energy under Grants DOE-ER-40682-143 and DEAC02-6CH03000. RAP also acknowledges support from the Foundational Questions Institute (fqxi.org) under grant RPFI-06-18, and funds from the University of California.

\newpage
 
\appendix
\section{On the Routhian formalism} 

Here we discuss some subtleties of the Routhian formalism, in particular its consistency with regards of the preservation upon time evolution of the SSC and the equivalence between imposing the SSC before or after obtaining the EOM.\\

Let us start with the expression in (\ref{dsu}). This is nothing but performing a field redefinition \cite{NRGR} (or coordinate transformation) given by $\delta x^\mu (\lambda) \sim S^{dc}u_c$ in the worldline action. Notice that it vanishes when one imposes the SSC. This implies that we could have indeed started using this other form if we wished, since the MP equations are also recovered. Notice that this extra acceleration dependent piece effectively entails adding a term\footnote{Recall from (\ref{up}) that $p^d \sim m \frac{u^d}{\sqrt{u^2}}$.} in the Routhian,
\beq S^{cd}u_c \frac{{\dot u}_d}{u^2}, \label{sav}\eeq 
and a modified gravity-spin interaction of the form \cite{yee}
\beq
-\frac{1}{2}\omega_\mu^{ab} {\cal S}_{ab} u^\mu \label{newcals},
\eeq
with
\beq
{\cal S}^{ab} = S^{ab} + \frac{u_c}{u^2}S^{c[a}u^{b]}\label{newsc}.
\eeq
 
The term in (\ref{sav}) becomes crucial, and generates a piece ${\dot u}_dS^{dc}u_c$ into the potential (recall ${\cal R} = -V$). It is simple to show that these extra terms do not effect the LO spin potentials, and that the 3PN results reported in this paper are also reproduced.
Notice we have now ${\cal S}^{ab}u_b=0$ algebraically. Written this way the SSC is manifestly preserved as we will now show. The algebra in terms of
 ${\cal S}^{ab}$ is that of (\ref{als}) with $\eta^{ab} \to \eta^{ab} - \frac{u^au^b}{u^2}$ as in (\ref{dirac3}). We can then show that 
\beq u_b\{ S^{ab},{\cal S}^{cd}\}=0. \label{calsu}\eeq  

From here we have, using (\ref{eom1}), 
\beq
\frac{d}{dt} (S^{ab}u_b)= u_b \{S^{ab}, {\cal R}_0 ({\cal S}^{ab})\}+{\dot u}_d\frac{u_c}{u^2}\{ S^{ab},S^{cd }\}u_b  + S^{ab}{\dot u}_b = 0,
\eeq
since
\beq
u_b \{S^{ab}, {\cal R}_0 ({\cal S}^{ab})\}=0\label{ros},
\eeq
where ${\cal R}_0$ stands for the Routhian without the acceleration dependent term. The same obviously follows from the Routhian in (\ref{actR}). The expression in (\ref{ros}) guarantees that higher dimensional operators written in terms of ${\cal S}^{ab}$ will preserve the SSC upon evolution.\\ 

Regarding the equivalence between imposing the SSC before or after obtaining the EOM,
we will show that the extra piece in the equations of motion in the Routhian formulation
exactly reproduces the Dirac bracket structure when imposing the SSC at the level of the
action. We start by noticing that the LO Dirac algebra of $S_{ab}$ agrees with that of Poisson algebra of  ${\cal S}^{ab}$. Since the only term in the Routhian which does not depend on ${\cal S}^{ab}$ is the one in (\ref{sav}), the above equivalence seems to rely on whether (\ref{sav}) can account for the extra pieces induced by the non--canonical algebra of (\ref{dirac1},\ref{dirac2},\ref{dirac3}). In what follows we show that is the case. The extra acceleration piece in the Routhian produces a term in the spin EOM given by (at LO ($a^0 \sim 0$) )
\beq
\{S^{ab},S^{cd}\}{\dot u_d} u_c \sim {\dot u}_c \left(S^{ac}u^b-S^{bc}u^a \right).\label{dotps} 
\eeq

 After imposing the covariant SSC (using Dirac brackets), $\frac{d}{dt}(S_{ab}u^b)=0$ leads to
\beq
u_b [S^{ab},H(x,p)]_{Db} = S^{ba}{\dot u}_b, 
\eeq
which implies
\beq
\label{A9}
\frac{dS^{ab}}{dt} = [S^{ab},H(x,p)]_{Db} = {\dot u}_c(S^{ac}u^b - S^{bc}u^a) + F^{ab},
 \eeq
with $F^{ab}=-F^{ba}$ and $F^{ab}u_b=0$. By comparison with (\ref{ros}) we immediately recognize that $F^{ab}$ is nothing but the Poisson bracket structure in $[~,~]_{Db}$. Therefore the extra piece 
in (\ref{A9}) comes from the non--canonical part of the Dirac bracket as we advertised above, which agrees with the acceleration dependent term  in (\ref{dotps}), and the equivalence follows.

\section{Feynman rules: spin--graviton vertex}

The spin-less part of the Feynman rules are identical to those in \cite{eih}.
The potential graviton propagator in the gauge of \cite{NRGR} is given by  
 \begin{equation}
\langle H_{\mu\nu} (x)H_{\alpha\beta}(0)\rangle= -iP_{\mu\nu;\alpha\beta}\int \frac{d^3k}{(2\pi)^3} \frac{1} {\bf
k^2}e^{-i {\bf k}\cdot {\bf x}} \delta(x_0)
\end{equation}
where 
\beq
P_{\mu\nu;\alpha\beta} = {1\over 2}\left[\eta_{\mu\alpha}
\eta_{\nu\beta} + \eta_{\mu\beta} \eta_{\nu\alpha} - \eta_{\mu\nu}
\eta_{\alpha\beta}\right].
\eeq
  The  mass vertices can be read off from $\cal R$ ,  
\begin{equation}
L_m=-\sum_{a=1,2}\frac{m_a}{m_p} \left[{1\over 2} H_{00} + H_{0i} v_{ai}  + {1\over 4} H_{00} {\bf v}^2_a + {1\over 2} H_{ij} 
v_{ai} v_{aj} \right]+ \ldots ,
\end{equation}
 where we have only included terms which are suppressed by $v^2$, as higher
 order terms will not contribute at 3PN once full diagrams are computed. In this expression, and from now on, we chose $\lambda=x^0=t$, and therefore $u^\mu = (1, \frac{d\bf x}{dt} \equiv {\bf v})$.
 The fields have arguments which are the wordline coordinates and   there is an implied affine parameter integral. Each diagram will contain an overall time integration which is dropped 
 when extracting the potential.\\ 
 
Following standard power counting procedures one arrives at  the scaling laws for the NRGR fields shown in table I. In the last column we have introduced
$m^2_p=\frac{1}{32\pi G_N}$ the Planck mass and $L=mvr$ the
angular momentum, with $v,r$ the relative velocity and orbit scale respectively. We will not consider radiation in this paper.
By  including the appropriate  spin vertices  
 the higher order radiation can be calculaled following the methodology
introduced in \cite{NRGR}. 

\begin{table}
\begin{eqnarray}
\nonumber
\begin{array}{c|c|c|c}
  {\bf k} & H^{\bf k}_{\mu\nu} &  m/m_p\\
\hline
   1/r      & r^2 v^{1/2} & \sqrt{L v}\\
\end{array}
\end{eqnarray}
\caption{NRGR power counting rules for potential modes.} \label{tab}
\end{table}
The non--linear graviton interactions are obtained from  expanding out the Einstein--Hilbert action \cite{NRGR}. The Feynman rule for the three graviton vertex has not been written down due to its length\footnote{In \cite{kol2}, it has been shown that these complicated diagrams can be eliminated
by a different choice of the metric}.
A Mathematica code showing how to include this vertex can be found at  \cite{workshop}.

 In order to obtain the spin--graviton vertex we also need to expand the metric in the weak gravity limit. In terms of the verbein we have
\begin{equation}
\label{ee}
e^a_{\mu}e^b_{\nu} ~\eta_{ab} = \eta_{\mu\nu} + \frac{H_{ \mu\nu}}{m_p}, ~ ~ e^a_{\mu} = \delta^a_{\mu} + \delta e^a_\mu \to  \delta e^a_\mu = \frac{1}{2m_p} H_\mu^a - \frac{1}{8m_p^2} H^a_\gamma H_\mu^\gamma+\ldots 
\end{equation}

Using (\ref{ee}) we can now expand the Ricci coefficients $\omega_\mu^{ab}$ in the weak gravity limit and extract the spin--graviton vertex rules \cite{nrgr3},
\begin{eqnarray}
L^{NRGR}_{1PN} &=& \frac{1}{2m_p}H_{i0,k}S^{ik},\label{sgnr1}\\
L^{NRGR}_{1.5PN} &=& \frac{1}{2m_p}\left(H_{ij,k}S^{ik}v^j + H_{00,k}S^{0k}\right),\label{sgnr15}\\
L^{NRGR}_{2PN} &=& \frac{1}{2m_p}\left(H_{0j,k}S^{0k}v^j +
H_{i0,0}S^{i0}\right)\nonumber
\\ &+& \frac{1}{4m^2_p}S^{ij}\left(H^{\lambda}_j H_{0\lambda,i} -
H^k_j H_{0i,k}\right).\label{sgnr2}
\end{eqnarray}
 
The appropriated scaling of each vertex is derived from the rules in (\ref{tab}). Furthermore, the components of the spin tensor scale as  $S^{0i} \sim v^j S^{ij}$, and $S \sim L v$ for maximally rotating compact bodies.

\end{document}